\documentclass[10pt,a4paper]{article}

\usepackage[utf8]{inputenc}
\usepackage{amsmath, amssymb, amsthm}
\usepackage{geometry}
\usepackage{setspace}
\usepackage{graphicx}
\usepackage{booktabs}
\usepackage[round, authoryear]{natbib}
\usepackage{xcolor}
\usepackage{authblk}
\usepackage{enumitem}
\usepackage{hyperref}
\usepackage[toc,page]{appendix}
\title{A joint meta-analysis framework for  the accuracy of two diagnostic tests accounting for varying study designs}

\date{}
\author[1]{Vera Hudak*}
\author[2]{Nicky J. Welton}
\author[2]{Efthymia Derezea}
\author[2]{Hayley E. Jones}

\affil[1]{School of Mathematics, University of Bristol, Bristol, UK}
\affil[2]{Population Health Sciences, Bristol Medical School, University of Bristol, Bristol, UK}

\vspace{1.5cm}

\affil[*]{Corresponding Author: vera.hudak@bristol.ac.uk}

\begin{document}

\maketitle

\begin{abstract}

Meta-analyses of the accuracy of two diagnostic tests typically assume tests are independent conditional on true disease status. This assumption is often unrealistic and violation leads to biased estimates of the accuracy of tests used in combination. Existing models accounting for conditional dependence require `joint classification' data (results for both tests and the `gold standard' on all participants) from all studies and/or suffer from computational instability. 

We propose a Bayesian hierarchical model for joint meta-analysis of the accuracy of two binary tests, modelling conditional dependence through study-specific log-odds ratios. The model accommodates studies that do not report joint classification data. We show how the model extends to accommodate data from varied study designs, including studies without a gold standard and studies with partial verification, without assuming imperfect reference standards are error-free. We demonstrate the framework with two example meta-analyses. 

Our modelling framework retains key features of standard diagnostic test accuracy meta-analysis methods, while allowing for conditional dependence. Ignoring conditional dependence yields biased joint accuracy estimates when conditional dependence is substantial. Our parametrisation maintains computational stability and accommodates data from varied study designs, without requiring an initial data imputation step or assuming error-free reference standards in all studies. 

    
\end{abstract} 

\newpage

\section{Introduction}

Meta-analysis of diagnostic test accuracy synthesises evidence on the accuracy of a diagnostic test across multiple studies \citep{Reitsma2005, Chu2006, Rutter2001}. In a typical diagnostic test accuracy study, the results of an index test (the test under evaluation) are compared to those of a reference standard, the best available method for determining the true disease status. Ideally the reference standard can be considered to be 100\% accurate, in which case we refer to it as a gold standard. The results of such studies are usually summarized in $2 \times 2$ contingency tables, which classify participants as true positives, false positives, false negatives and true negatives. From these tables, sensitivity and specificity of the index test can be estimated in each study as the proportion of diseased individuals correctly identified by the test and the proportion of disease-free individuals correctly identified by the test, respectively. To synthesise evidence across studies, bivariate random effects meta-analysis of sensitivity and specificity is widely recommended, which accounts for the trade-off between these measures and allows for variation in test performance across studies. 

Several approaches have been proposed to jointly meta-analyse the accuracy of two or more index tests \citep{Ma2018, Menten2015, Lian2019, Nyaga2016, Nikoloulopoulos2019, Hoyer2018}, many of which are extensions of the bivariate random effects model. These more complex synthesis methods are needed to address several key clinical questions: identifying which of several available tests is most accurate for use as a stand-alone diagnostic tool, and evaluating the accuracy of tests when used in combination according to different decision rules. However, the majority of these methods assume that the results of the index tests are conditionally independent given the true disease status. This assumption is often unrealistic, particularly when tests that are based upon a common biological mechanism are applied to the same individuals \citep{Vacek1985}, or when disease occurs in varying degrees of severity and tests are more likely to yield positive results in subjects with severe disease \citep{pepe2007}. The assumption of conditional independence is especially problematic when interest lies in the accuracy of tests used in combination. This is often the case, since in clinical practice tests are often used in sequence or in combination \citep{Heimbach2018, NICE2016, WHO2021Cervical}. Failure to account for conditional dependence leads to biased estimates of the accuracy of tests used in combination, and invalid conclusions \citep{vanWalraven2009}. Furthermore, it may also lead to a loss of precision in comparative test accuracy estimates \citep{trikalinos_2014, McBride2025}.

Only a few meta-analysis models have been proposed that can be used to produce summary estimates of the accuracy of two index tests used in combination or sequence, accounting for conditional dependencies \cite{trikalinos_2014, Novielli2013, Nikoloulopoulos2024, McBride2025}. The model proposed by \citet{trikalinos_2014} assumes multinomial likelihoods for data from joint classification tables and a 6-dimensional multivariate normal distribution for the joint and individual true positive and false positive rates on the logit scale across studies. However, joint rates are naturally constrained by the individual sensitivities and specificities, therefore their logit-transformed values do not behave as unconstrained parameters in a random-effects framework, which can lead to significant estimation issues (see Discussion). Additionally, this model requires data from full $ 2 \times 4 $ joint classification tables \citep{Yang2021} from every study, reporting cross-classified results of the two index tests against each other for participants in both the diseased and disease-free groups. In practice, these data are often unavailable in published reports of diagnostic test accuracy studies. Trikalinos et al. rely on imputation, as an initial stage prior to meta-analysis, when only $2 \times 2$ contingency tables are reported separately for each index test. 

More recently, \citet{Nikoloulopoulos2024} and \citet{McBride2025} used copula functions to jointly meta-analyse the accuracy of two tests while accounting for conditional dependence. Nikoloulopoulos adapted the model of \citet{trikalinos_2014} by maintaining the multinomial likelihood for joint classification tables but replacing the multivariate normal random effects with a D-vine copula representation, with either normal or beta margins. McBride proposed a 6-dimensional multivariate normal likelihood to account for between-study variation, and a trivariate copula to link the results of the two tests within each study. Using copulas allows for the modelling of complex dependence shapes, such as tail dependence where tests agree more strongly at extreme accuracy levels. However, all of these approaches assume a random effects distribution for the joint rates across studies, and thus do not resolve the constraint issues discussed above. Furthermore, like the Trikalinos et al. model, they each require data from full joint classification tables from every study for estimation.

Finally, \citet{Novielli2013} modelled conditional dependence using a stratification approach, in which one test acts as a categorical risk stratifier for the other. One potential disadvantage of this model is that it is not symmetrical, such that results might differ depending on which test is used as the risk stratifier, which in the case of two binary tests can often be arbitrary. Furthermore, when studies fail to report stratified results, the model relies on stochastically predicting the missing proportions of the study population.

As noted above, many primary studies do not report full joint classification tables, instead reporting only separate $2 \times 2$ contingency tables for each index test. Beyond such data reporting issues, the synthesis of diagnostic accuracy is frequently complicated by heterogeneity in study designs and the presence of verification bias. Some studies might have evaluated the two index tests on different populations, only evaluated one of the two tests, or have evaluated both tests but with missing results on one test for some individuals. Some studies might not have used the preferred reference standard -- instead, perhaps evaluating one of the index tests against the other, treating the second as if it were a gold standard. In other studies, testing with the gold standard might be conditional on the result of one or both index tests.

In this paper, we propose a general modelling framework for the joint meta-analysis of the accuracy of two diagnostic tests. Our model overcomes the limitations associated with modelling joint true positive and joint false positive rates as random effects across studies, by modelling conditional dependence between tests through log-odds ratios, while also treating both tests equally (i.e. being symmetrical). Since log-odds ratios are mathematically unconstrained, this parametrisation avoids the boundary issues inherent in modelling joint rates, improving computational stability. Furthermore, while our approach requires some studies with joint classification tables to estimate within-study dependence, it allows for the direct inclusion of $2 \times 2$ contingency tables where joint data is not available, without relying on data imputation. Additionally, our framework naturally extends to include studies with alternative designs, by accommodating a variety of study-specific likelihood functions to match the observed data structure of each study.

The rest of this document is organized as follows: we begin by introducing two case studies to motivate the proposed modelling framework (Sections 2-3). We then describe the joint meta-analysis framework, parametrized via log-odds ratios, assuming initially that full joint classification tables are reported by every study (Section \ref{sec: jont MA framework}). We then extend this framework to include studies with varied designs and reporting (Section \ref{sec: varied designs}), followed by a description of inference and estimation procedures. Finally, we report results for both case studies (Section 5) and discuss our findings (Section 6). 

\section{Case Study 1: Accuracy of Second-Trimester Ultrasound \\ Markers in Screening for Down Syndrome}

Our first case study is from a systematic review of the  accuracy of second-trimester ultrasound markers in detecting liveborn infants with trisomy 21 (Down syndrome) \citep{smithbindman2001}. Following \citet{trikalinos_2014}, we focus on a subset of the full data set, describing the accuracy of two ultrasonographic markers, shortened humerus and shortened femur, as index tests. This data set was previously used by \citet{trikalinos_2014} and \citet{Nikoloulopoulos2024} to demonstrate their methods. The reference standard was cytogenetic karyotyping, based on specimens obtained through invasive procedures such as amniocentesis or chorionic villus sampling, which we will treat as a gold standard. The classification of study designs and reported data structures used here is based on the information presented in \citet{trikalinos_2014}.

\citet{Yang2021} classifies study designs for comparative diagnostic test accuracy into three primary categories: fully paired (where all participants receive both index tests), partially paired (where a random or non-random subset of participants does not receive both tests), and unpaired (where each participant receives only one of the index tests according to a random or non-random allocation). In this data set, there are 11 studies in total, and all studies evaluated both markers using paired designs. Eight studies \citep{benacerraf1991, benacerraf1992, benacerraf1994, biagiotti1994, nyberg1993, nyberg1998, rodis1991, vintzileos1996} had fully paired design, assessing the accuracy of both markers in every participant. However, the reporting of data from these studies was varied, leading to different observed data structures. In the remaining three studies \citep{bromley1997, johnson1995, lockwood1993}, humeral measurements were missing for a subset of participants. We will treat these three studies as having a partially paired design with missing humerus data for a random subset of participants, although the aim was likely a fully paired design. We will assume that missing humerus measurements were missing completely at random (MCAR), due to technical issues or study-specific data collection practices rather than factors related to disease status or test outcomes. The different study designs and reporting procedures are summarised below, and the full dataset can be found in Table \ref{tab:ultrasound_data} in Appendix \ref{appendix: data for case studies}.

\begin{enumerate}[label=(\alph*)]
\item 5 studies \citep{benacerraf1991, benacerraf1992, benacerraf1994, biagiotti1994, nyberg1993} had a fully paired design and reported $2\times4$ joint classification tables, with cross-classified results on shortened humerus, shortened femur and the reference standard.
\item 2 studies \citep{nyberg1998, rodis1991} had a fully paired design but reported $2 \times 2$ contingency tables for each marker separately, without cross-classified data that would allow estimation of joint test accuracy.
\item 1 study \citep{vintzileos1996} had a fully paired design and reported a joint classification for the diseased group only: cross-classified results for shortened femur and humerus were reported in infants with trisomy 21 but, in the disease-free group, each marker was reported separately without joint classification.
\item 3 studies \citep{bromley1997, johnson1995, lockwood1993} had a partially paired design with a randomly missing subset, where a (assumed random) subset of participants' humerus length was not reported. Data was reported for each marker separately. 
\end{enumerate}

The original systematic review and meta-analysis by Smith-Bindman et al. synthesized the data for each marker separately, focusing on the accuracy of each marker individually rather than estimating joint accuracy. Conditional dependence between shortened humerus and shortened femur was therefore not considered. While the joint meta-analyses of the two markers by \citet{trikalinos_2014} and \citet{Nikoloulopoulos2024} accounted for potential conditional dependence between the two markers, both approaches relied on an initial step, prior to meta-analysis, where missing cross-classified data were imputed, which may lead to overestimating the precision of the summary estimates. The approach proposed in this paper eliminates the need for this data imputation step.

\section{Case Study 2: Accuracy of McIsaac Clinical Prediction Rule in Identifying Group A Streptococcus}

In a systematic review of the accuracy of the McIsaac clinical prediction rule in identifying group A streptococcus infection in patients with pharyngitis \citep{kanagasabai2024}, the included studies varied in both design and reporting. Eight studies were included in the meta-analysis \citep{abd_el_ghany2015, cohen2012, ezike2005, felsenstein2014, palla2012, edmonson2005, lindgren2016, nishiyama2018}. While the researchers considered throat culture to be the best available reference standard (and we will treat it as the gold standard), this was not used consistently in all studies. In some studies, a rapid antigen detection test (RADT) served as an imperfect reference standard, either for all study participants or with only those testing positive on the RADT undergoing throat culture. Although the intention of the systematic review was to evaluate the accuracy of the McIsaac prediction rule only, a subset of studies additionally evaluated a RADT as a second index test. Differences in verification procedures and in how test results were reported led to variation in the available data structures, which we detail below. The full data extracted from these studies -- using a McIsaac score threshold of 3+ to define a positive result -- are presented in Table \ref{tab:mcisaac_data} in Appendix \ref{appendix: data for case studies}.

\begin{enumerate}[label=(\alph*)]
    \item 3 studies had a fully paired design, where all participants underwent McIsaac, RADT, and throat culture; however, only 2 \citep{cohen2012, ezike2005} of these reported $2\times4$ joint classification tables, while the third \citep{abd_el_ghany2015} reported $2\times2$ tables of McIsaac vs. throat culture, RADT vs. throat culture and McIsaac vs. RADT, but not the cross verified results.
    \item 1 study \citep{palla2012} was a single test accuracy study, where McIsaac was evaluated against throat culture only.
    \item 1 study \citep{felsenstein2014} had a partially paired design, where all participants received RADT and throat culture, but only those testing positive on either of these were subsequently assessed with the McIsaac score. The missing test results therefore cannot be assumed to be missing at random. Data was reported for RADT vs. throat culture in the full population, and for McIsaac vs. throat culture and McIsaac vs. RADT in the restricted population. 
    \item 1 study \citep{nishiyama2018} had an imperfect reference standard only: McIsaac was evaluated only against RADT.
    \item 2 studies \citep{edmonson2005, lindgren2016} had what we will refer to as a `check the negatives` design \citep{hudak2026quantifying}. Here, McIsaac and RADT were given to all participants, but only RADT negatives were further tested with throat culture. \citet{edmonson2005} reported cross-classified data for those undergoing the gold standard (RADT negatives) and only McIsaac vs RADT for those untested on gold standard (RADT positives). \citet{lindgren2016} reported McIsaac results against a composite reference standard combining RADT and throat culture. 
\end{enumerate}

For the original meta-analysis of the accuracy of the McIsaac rule, data from all eight studies were included \citep{kanagasabai2024}. McIsaac versus throat culture data were used where available (five studies). Where this was not available, McIsaac was compared against the best available reference standard for each study: i.e. RADT in one study and a composite reference standard of RADT and throat culture in two studies. As standard meta-analysis methods were applied, each  reference standard was implicitly assumed to be 100\% sensitive and 100\% specific. 

Note that, as the focus of \citet{kanagasabai2024} was on pooling estimates of the accuracy of the McIsaac prediction rule only -- with RADT used in some studies to define the (assumed error-free) reference standard, rather than as a second index test -- consideration of conditional dependence was not relevant. Furthermore, due to this focus, the review was restricted to studies evaluating McIsaac (either alone or paired with RADT), while studies evaluating RADT alone were ineligible for inclusion. In contrast, in this paper we treat both McIsaac and RADT as index tests, assessing their performance individually as well as when used in combination. This joint framework additionally allows us to estimate and allow for the conditional dependence between these tests and incorporate study-specific designs directly. 

\section{Methods}

In this section, we describe a Bayesian hierarchical model for the joint meta-analysis of the accuracy of two binary diagnostic tests, indexed by $k = \{1, 2\}$, from $I$ studies. Our approach incorporates conditional dependence between the tests, modelled through odds ratios. We begin by describing the model for the situation in which we have data from $2\times4$ joint classification tables on the two tests compared with a gold standard from every study. We then extend this framework to the varied study designs and reporting methods observed in our motivating examples.

\subsection{Joint Meta-Analysis Framework} \label{sec: jont MA framework}

Let Tests 1 and 2 denote the index tests, denoted by $k \in \{1, 2\}$, evaluated across a set of studies $i = 1, 2, \ldots, I$. For each study $i$ and test $k$, let $Se_{ki}$ denote the true sensitivity (also referred to as the true positive rate, $TPR_{ki}$) and $Sp_{ki}$ denote the true specificity, with $FPR_{ki} = 1 - Sp_{ki}$ representing the corresponding false positive rate. Furthermore, let $\pi_i$ be the prevalence of the condition of interest in the $i$-th study population. To capture possible conditional dependence between Tests 1 and 2, we introduce parameters $c_{0i}$ and $c_{1i}$, representing the covariance between the two tests in the disease-free and diseased populations in study $i$, respectively.

Let $t_{1}, t_{2}, d$ represent binary results for an individual on Test 1, Test 2, and the gold standard, respectively, where $t_{1}, t_{2}, d \in \{0,1\}$ correspond to negative (0) and positive (1) test results. The probability of a randomly selected individual having test results $(t_{1}, t_{2}, d)$ in study $i$, denoted $p_{t_1t_2di}$, is given in Table \ref{tab: fully paired design} \citep{Vacek1985}. Furthermore, let $y_{t_1t_2di}$ denote the observed number of individuals in study $i$ with test results $(t_{1}, t_{2}, d)$.

\begin{table}[htbp]
\centering
\caption{Underlying probabilities of every combination of test results when Test 1, Test 2, and the gold standard (GS) are carried out on all participants in study $i$, and the corresponding observed counts.}
\begin{tabular}{cccccc}
\toprule
\textbf{Test 1} & \textbf{Test 2} & \textbf{GS} & $p_{t_1 t_2 di}$ & \textbf{Probability} & \textbf{Observed Counts}\\
\midrule
- & - & - & $p_{000i}$ & $(1-\pi_i)(Sp_{1i} Sp_{2i} + c_{0,i})$ & $y_{000i}$ \\
+ & - & - & $p_{100i}$ & $(1-\pi_i)((1-Sp_{1i}) Sp_{2i} - c_{0i})$ & $y_{100i}$\\
- & + & - & $p_{010i}$ & $(1-\pi_i)(Sp_{1i} (1-Sp_{2i}) - c_{0i})$ & $y_{010i}$ \\
+ & + & - & $p_{110i}$ & $(1-\pi_i)((1-Sp_{1i})(1-Sp_{2i}) + c_{0i})$ & $y_{110i}$ \\
- & - & + & $p_{001i}$ & $\pi_i((1-Se_{1i})(1-Se_{2i}) + c_{1i})$ & $y_{001i}$\\
+ & - & + & $p_{101i}$ & $\pi_i(Se_{1i}(1-Se_{2i}) - c_{1i})$ & $y_{101i}$\\
- & + & + & $p_{011i}$ & $\pi_i((1-Se_{1i}) Se_{2i} - c_{1i})$ & $y_{011i}$\\
+ & + & + & $p_{111i}$ & $\pi_i(Se_{1i} Se_{2i} + c_{1i})$ & $y_{111i}$\\
\bottomrule
\end{tabular}
\label{tab: fully paired design}
\end{table}

Suppose we observe fully cross-classified data on Tests 1, 2 and the gold standard for every study $i$, that is, we observe each $y_{t_1t_2di}$ from Table \ref{tab: fully paired design}. We model this using two 4-dimensional multinomial distributions, one for the diseased population ($d = 1$) and one for the disease-free population ($d = 0$): 

\begin{align} 
\mathbf{y}_{1i} &\sim \text{Multinomial}(N_{1i}, \mathbf{p}_{1i}) \\ 
\mathbf{y}_{0i} &\sim \text{Multinomial}(N_{0i}, \mathbf{p}_{0i}),
\end{align}

where $\mathbf{y}_{di} = (y_{00di}, y_{10di}, y_{01di}, y_{11di})$ denotes the vector of counts for test results ($t_1, t_2$), $\mathbf{p}_{di} = (p_{00di}, p_{10di}, p_{01di}, p_{11di})$ the corresponding joint probabilities as defined in Table \ref{tab: fully paired design} and $N_{di}$ the total number of participants in study $i$ with true disease status $d$.

\subsubsection{Modelling Joint Rates via Log-Odds Ratios}
\label{sec: joint sens and spec log or}

A key challenge in jointly meta-analysing of the accuracy of two diagnostic tests is the between-study modelling of the covariance parameters, $c_{0i}$ and $c_{1i}$. The feasible ranges for these parameters differ across studies because the lower and upper bounds are functions of the study-specific sensitivities and specificities, as follows \citep{Dendukuri2001}:

\begin{align*}
    (Se_{1i}-1)(1-Se_{2i}) & \leq c_{1i} \leq \min(Se_{1i}, Se_{2i}) - Se_{1i}Se_{2,i}, \\
    (Sp_{1i}-1)(1-Sp_{2i}) & \leq c_{0i} \leq \min(Sp_{1i}, Sp_{2i}) - Sp_{1i}Sp_{2i}.
\end{align*}

In a single study setting, these constraints can be enforced by using uniform prior distributions, with the bounds defined as above, or following \citet{Dendukuri2001}, using a generalized beta distribution for each covariance parameter. Similarly, in latent class meta-analysis of test accuracy (with 2 index tests but no gold standard), covariance parameters have been treated as independent across studies, with study-specific uniform prior distributions \citep{CochraneHandbook2023}. While this ensures the dependence remains within the feasible range for each study, it treats conditional dependence as a study specific parameter rather than modelling a common between-study dependence structure.

It is challenging to specify common between-studies distributions for these parameters, facilitating borrowing of information across studies and production of summary estimates of joint test accuracy, while also ensuring that the study-specific bounds are enforced. 

\citet{trikalinos_2014} proposed parametrising in terms of the joint true positive rate (joint TPR), $p_{111i}$, and joint false positive rate (joint FPR), $p_{110i}$, but these quantities are also naturally constrained by the individual TPRs and FPRs ($0 \leq p_{111i} \leq min(TPR_{1i}, TPR_{2i})$ and $0 \leq p_{110i} \leq min(FPR_{1i}, FPR_{2i})$).  Alternatively, \citet{Chu2009} assumed constant Pearson correlations across studies, constrained to the tightest constraints across each individual study. However, like joint rates and the covariance parameters, Pearson correlations are naturally constrained by study-specific sensitivities and specificities. Consequently, they do not serve as intuitive parameters for random or common effects.

Instead of assuming that the joint rates, covariance parameters, or Pearson correlations are random effects, we propose modelling within-study dependencies using log-odds ratios. Log-odds ratios are unconstrained and are a more natural scale to place random effects on. We present the derivation of this approach below, omitting the study index $i$ for simplicity. 

First focusing on the diseased population, the odds ratio representing the association between Tests 1 and 2 is: 

\begin{align*}
    OR_1 = \frac{p_{111}p_{001}}{p_{011}p_{101}},
\end{align*}

which we can re-express as follows:

\begin{align*}
    OR_1 = \frac{p_{111}(1+p_{111}-TPR_1-TPR_2)}{(TPR_1-p_{111})(TPR_2-p_{111})}.
\end{align*}

Rearranging and then solving for $p_{111}$ (joint TPR) gives:

\begin{align*}
    p_{111} = 
\begin{cases} 
      \frac{((OR_1-1)(TPR_1+TPR_2)+1) \pm \sqrt{((OR_1-1)(TPR_1+TPR_2)+1)^2 - 4OR_1(OR_1-1)TPR_1TPR_2}}{2(OR_1-1)} & OR_1 \neq 1 \\
      TPR_1TPR_2 & OR_1=1 
\end{cases}
\end{align*}

Similarly, the odds ratio representing the association between Tests 1 and 2 in the disease-free population is:
\begin{align*}
    OR_0 = \frac{p_{110}p_{000}}{p_{010}p_{100}},
\end{align*}
and we can similarly derive an expression for $p_{110}$ (joint FPR) in terms of $OR_0$ and the FPRs:

\begin{align*}
    p_{110} = 
\begin{cases} 
      \frac{((OR_0-1)(FPR_1+FPR_2)+1) \pm \sqrt{((OR_0-1)(FPR_1+FPR_2)+1)^2 - 4OR_0(OR_0-1)FPR_1 FPR_2}}{2(OR_0-1)} & OR_0 \neq 1 \\
      FPR_1 FPR_2 & OR_0=1 
\end{cases}
\end{align*}

For both $p_{111}$ and $p_{110}$, we find that only the negative root yields a valid solution. The derivation can be found in Appendix \ref{appendix: negative root}.

\subsubsection{Between-Studies Model}
\label{sec: model}

As in standard bivariate random effects meta-analysis of test accuracy, we model individual TPRs and FPRs on the logit (log-odds) scale across studies. To incorporate the conditional dependence between tests within the disease-free and diseased populations, we additionally model study-specific log-odds ratios, log($OR_{0i}$) and log($OR_{1i}$), where $OR_{0i}$ and $OR_{1i}$ are defined as  in Section \ref{sec: joint sens and spec log or}. To model the joint distribution and account for between-study heterogeneity, we assume that the vector of logit-transformed TPRs, FPRs, and log-odds ratios follows a 6-dimensional multivariate normal distribution:

\begin{align*}
    \begin{pmatrix}
\operatorname{logit}(TPR_{1i}) \\
\operatorname{logit}(TPR_{2i}) \\
\log(OR_{1i}) \\
\operatorname{logit}(FPR_{1i}) \\
\operatorname{logit}(FPR_{2i}) \\
\log(OR_{0i})
\end{pmatrix}
\sim \operatorname{N}
\left(
\boldsymbol{\mu}
, \mathit{\Sigma}
\right),
\end{align*}

where the mean vector $\boldsymbol{\mu} = (\mu_{11}, \mu_{21}, \eta_1, \mu_{10}, \mu_{20}, \eta_0)$ contains the overall average value for each parameter on the transformed scale. Specifically, $\mu_{11}$ and $\mu_{21}$ represent the pooled logit TPRs for Test 1 and Test 2, respectively, while $\mu_{10}$ and $\mu_{20}$ represent the pooled logit FPRs. The parameters $\eta_{1}$ and $\eta_{0}$ represent the mean log-odds ratios, capturing the average conditional dependence between the two tests in the diseased and disease free populations. 

These pooled estimates can be transformed back to the probability scale to obtain summary estimates for the true and false positive rates and odds ratios. Let us denote these summary estimates by $TPR_{1}^s, TPR_{2}^s, OR_{1}^s, FPR_{1}^s, FPR_{2}^s$ and $OR_{0}^s$. Then for example, the summary TPR for Test 1 is estimated as $TPR_{1}^s = \frac{exp(\mu_{11})}{1+exp(\mu_{11})}$, while the summary odds ratio in the diseased population is $OR_{1}^s = exp(\eta_1)$. We can find $TPR_{2}^s, FPR_{1}^s, FPR_{2}^s$ and $OR_0^s$ similarly. Using these summary estimates, we can derive any joint accuracy measure of interest. For example, the summary probability of testing positive on both tests among diseased individuals, is estimated as:

$$\text{Joint TPR} =
\begin{cases}
\frac{1 + ({TPR}_1^s + {TPR}_2^s)({OR}_1^s - 1) - \sqrt{[1 + ({TPR}_1^s + {TPR}_2^s)({OR}_1^s - 1)]^2 - 4{OR}_1^s({OR}_1^s - 1){TPR}_1^s{TPR}_2^s}}{2({OR}_1^s - 1)} & \text{if } {OR}_1^s \neq 1 \\
{TPR}_1^s{TPR}_2^s & \text{if } {OR}_1^s = 1
\end{cases}$$

We can find the joint FPR similarly, using the posterior distributions of $FPR_1, FPR_2$ and $OR_0$. 

The $6 \times 6$ variance-covariance matrix $\mathit{\Sigma}$ allows for  between-study heterogeneity for each parameter, as well as the correlations between these parameters across studies.

\subsection{Extension to Varied Study Designs and Reporting} \label{sec: varied designs}

The multinomial likelihoods described above assume that every participant undergoes both tests and is verified by the gold standard, and that complete cross-classified data are available. However, in systematic reviews, studies often employ alternative designs or report data incompletely. We extend our framework to accommodate the varied study designs and reporting methods observed in our motivating examples. We assume that for every study $i$, the latent parameters $TPR_{k, i}, FPR_{k, i}, OR_{0, i}$ and $OR_{1, i}$ are drawn from the same 6-dimensional multivariate normal distribution as in Section \ref{sec: model}. The variation across designs is handled by modifying the likelihood function to match the observed data structure as described below.

Across all the designs and reporting variations detailed in this section, while the likelihood functions are tailored to the specific observed data structure of each study, the latent parameters ($TPR_{k,i}, FPR_{k,i}$, $OR_1,i$ and $OR_0,i$) for each study $i$ are drawn from the same 6-dimensional multivariate normal distribution defined in Section \ref{sec: model}. This allows the model to share information across studies $i = 1, \ldots, I$ regardless of their specific reporting format or verification design.

\subsubsection{Complete Verification}
\label{sec: complete verification}

We start by describing the likelihoods observed in the ultrasound marker example, where all participants undergo the gold standard and missing results on any index test are assumed MCAR. For studies reporting cross-classified data, the likelihoods are as defined by Equations 1 and 2. 

\vspace{0.4cm}

\noindent \textbf{Incomplete reporting in a fully paired design and unpaired designs}: Where a study evaluates both index tests on separate people, or on the same people but reports each one separately, i.e.  Test 1 vs. gold standard, Test 2 vs. gold standard (like in studies \citet{nyberg1998} and \citet{rodis1991} from Case Study 1), then it contributes no information on the presence or extent of conditional dependence. 

In this case, we model the data as four binomial distributions (two per test) conditional on disease status. Let $y_{1.1,} = y_{111i} + y_{101i}$ and $y_{0.1i} = y_{011i} + y_{001i}$ denote the the number of true positives and false positives on Test 1 in study $i$, respectively, and $y_{.11i} = y_{111i} + y_{011i}$ and $y_{.01i} = y_{101i} + y_{001i}$ denote the the number of true positives and false positives on Test 2 in study $i$, respectively. Then:

\begin{align} y_{1.1i} &\sim \text{Binomial}(N_{1i}, TPR_{1i})  \\
y_{0.1i} &\sim \text{Binomial}(N_{0i}, FPR_{1i}) \\
 y_{.11i} &\sim \text{Binomial}(N_{1i}, TPR_{2i}) \\
y_{.01i} &\sim \text{Binomial}(N_{0i}, FPR_{2i}). 
\end{align}

Occasionally, studies may also report the $2 \times 2$ contingency table of Test 1 vs. Test 2 results (as in the study \citet{abd_el_ghany2015} from Case Study 2). Notably, this third table does not provide entirely new information, as the total number of positives on Test 1 and Test 2 are already fixed by the data from the Test 1 vs. gold standard and Test 2 vs. gold standard tables. Therefore this table provides only one unique piece of information, which is the overall degree of overlap between the two tests, across the two disease classes, which might provide some information on test dependence. To incorporate this data without redundancy, we model the count of individuals testing positive on both tests in study $i$, denoted $y_{11\cdot,i} = y_{110,i} + y_{111,i}$ as: 
\begin{equation*}
y_{11\cdot i} \sim \text{Binomial}(N_i, p_{11\cdot i})
\end{equation*}

where $N_i = N_{1i} + N_{0i}$ is the total sample size of study $i$ and $p_{11\cdot,i} = p_{111,i} + p_{110,i}$ is the overall probability of a joint positive results, across both diseased and disease-free populations.

Additionally, some studies might report partially cross-classified data, where the reporting format differs between the diseased and disease-free populations. In this case, the study contributes a multinomial likelihood in one disease state (where cross-classified data is reported) and two binomial likelihoods in the other disease state (where data is reported separately for tests). For example, in \citet{vintzileos1996} from Case Study 1, cross-classified data were reported for the diseased population, but each marker was reported separately for the disease-free population. The data from this study contribute Equation 1 in the diseased state and Equations 4 and 6 in the disease-free state. 

\vspace{0.4cm}
\noindent \textbf{Single test accuracy studies}
If the decision is made to include single test accuracy studies in the meta-analysis (see Discussion), these can easily be incorporated into the framework. Each contributes a single pair of binomial likelihoods, as defined by Equations 3 and 4 if the study evaluates Test 1, or Equations 5 and 6 if it evaluates Test 2.

\vspace{0.4cm}

\noindent \textbf{Partially paired design  with a randomly missing subset}: We next consider a partially paired design with a randomly missing subset, where all participants are tested with the gold standard and Test 1, but an (assumed MCAR) subset of Test 2 results are missing. Therefore we only have results for Test 2 in a subset $S$ of the full population. This is the design of three studies from the ultrasound marker example\citep{bromley1997, johnson1995, lockwood1993}. If the fully cross-classified data for the individuals in subset $S$ were available, this should be modelled as per Table \ref{tab: fully paired design}. The additional test results for the individuals not in $S$ (who are missing Test 2 results) can also be incorporated into the model through additional binomial likelihoods for Test 1, using the total number of diseased and disease-free individuals who did not receive Test 2 as denominators.

For the three examples from our Case Study 1, no cross-classified results were reported. In the absence of any cross-classified data, the study contributes independent binomial likelihoods for each test as per Equations 3–6, using the full population as the denominator for Test 1 and the restricted subset $S$ as the denominator for Test 2.

\subsubsection{Alternative Verification Methods}

In Section \ref{sec: complete verification}, we described likelihoods for study designs in which every participant is verified by a gold standard and any missing data on Tests 1 or 2 is MCAR. We now consider studies that employ alternative verification strategies, where not every participant undergoes the gold standard. Testing with the gold standard may be conditional on the results on Test 1 and/or Test 2, in which case missing data is not missing at random (NMAR). We need appropriate likelihood modelling to allow for these designs, to avoid verification bias.

For example, in Case Study 2, in one study only those testing positive on the gold standard or the first index test received the second one. In another one, only an imperfect reference standard was used for verification, and in two studies, gold standard testing was conditional on the outcome of the second index test. We do not provide a comprehensive list of likelihoods for all possible study designs, but describe the likelihoods for these studies from Case Study 2 below, to demonstrate the concepts.  

\vspace{0.4cm}

\noindent \textbf{Imperfect reference standard study}: Under this design, Test 1 outcomes are verified by Test 2 only. The observed probabilities are obtained by marginalising over the unobserved gold standard result. This latent class approach is common in diagnostic test research where the gold standard is missing \citep{Vacek1985, Hui1980, Johnson2001}. We model the observed counts as a single 4-dimensional multinomial distribution:

\begin{align*}
    \mathbf{y}{\cdot}_i \sim \text{Multinomial}(N_i, \mathbf{p}{\cdot}_i),
\end{align*}

where $\mathbf{y}{\cdot}_i = (y_{00\cdot i}, y_{01\cdot i}, y_{10\cdot i}, y_{11\cdot} i)$ denotes the vector of joint counts for Test 1 and Test 2 in study $i$ in both diseased states, with $y_{t_1t_2\cdot,i} = y_{t_1t_20 i} + y_{t_1t_21 i}$, and $\mathbf{p}_i{\cdot} = (p_{00\cdot i}, p_{01\cdot i}, p_{10\cdot i}, p_{11\cdot i})$ are the corresponding cell probabilities, with $p_{t_1t_2\cdot i} = p_{t_1t_20 i} + p_{t_1t_21 i}$.

\vspace{0.4cm}

\noindent \textbf{Partially paired design with a non-random subset}: We next consider a partially paired design with a non-random subset, where all participants are tested with the gold standard and Test 1, but only those testing positive on at least one of these tests undergoes Test 2. This is the design of \citet{felsenstein2014} from Case Study 2.

We first consider the case where fully cross-classified results are reported in such a study. In this design, we observe almost all cells in Table \ref{tab: fully paired design}, but as Test 1 negatives in the disease-free population did not receive Test 2, the two corresponding cells are collapsed into a single observed count ($y_{0\cdot 0 i} = y_{010i} + y_{000i}$). This results in seven unique observed outcome combinations across the diseased and disease-free populations, which can be modelled with a 7-dimensional multinomial distribution:

\begin{align*} \mathbf{y}_{pp_i} \sim \text{Multinomial}(N_i, \mathbf{p}_{pp_i}),
\end{align*}

where $\mathbf{y}_{pp_i} = (y_{100i}, y_{110i}, y_{001i}, y_{101i}, y_{011i}, y_{111i}, y_{0\cdot0,i})$ are the counts in study $i$, and \newline $\mathbf{p}_{pp_i} = (p_{100i}, p_{110i}, p_{001i}, p_{101,}, p_{011i}, p_{111i}, p_{0\cdot0i})$, with $p_{0\cdot0,i} = p_{000,i} + p_{010,i}$, are the corresponding probabilities.

However, in the Felsenstein et al. study, cross-classified results are not reported. Instead, they report Test 1 vs gold standard in the whole population, and Test 2 vs Test 1 and Test 2 vs gold standard in the restricted population of those who tested positive on the gold standard or Test 1.  

As the Test 1 vs. gold standard table is reported for the whole population, and data can be modelled with two binomials as in Equations 3 and 4. Unlike in Section \ref{sec: complete verification} where data was missing MCAR, the subset of individuals undergoing Test 2 is not completely random. The probability of belonging to the tested subset in study $i$ is given by:

\begin{align*}
    \mathbb{P}(S_i) = \mathbb{P}(d = 1 \text{ or }t_1 = 1)_i = \pi_i + (1-\pi_i)FPR_{1,i},
\end{align*}

that is, all diseased individuals plus the non-diseased who test positive on Test 1. The number of true positives on Test 2, $y_{.11}$, can be modelled using a binomial distribution:

\begin{equation*}
y_{.11} \sim \text{Binomial}(N_{1|S_i}, TPR_{2i}), 
\end{equation*}

where $N_{1|S_i}$ is the number of diseased individuals in subset $S$ of study $i$. Note that since all participants undergo the gold standard, $N_{1|S_i}$ and $N_{0|S_i}$ are observed constants. Theoretically, the entire diseased population is included in $S$. However, to account for incidental missing data or study-specific exclusions, we condition on  $N_{1|S,i}$ rather than the total diseased population. For the disease-free population, inclusion in $S$ is conditional on having tested positive on Test 1. Therefore, the number of false positives on Test 2, $y_{.01}$, is modelled using a binomial distribution: 

\begin{equation*} 
y_{.11} \sim \text{Binomial}(N_{0|S_i}, \theta_i),
\end{equation*}

where $N_{0|S_i}$ is the number of disease-free participants who tested positive on Test 1 (the false positives on Test 1). The parameter $\theta_i$ is the probability of testing positive on Test 2 given a positive result on Test 1 along the disease-free population, which is the ratio of the joint false positive rate to the false positive rate of Test 1: 

\begin{equation*} \theta_i = \frac{\mathbb{P}(T_1=1, T_2=1 | D=0)_i}{\mathbb{P}(T_1=1 | D=0)_i} = \frac{p_{110i}}{(1-\pi_i)FPR_{1i}}.
\end{equation*}

If data on Test 1 vs. Test 2 are also reported (as in \citet{felsenstein2014}), the marginal totals for both tests within the restricted population are fixed by the data from the Test 1 vs. gold standard and Test 2 vs. gold standard tables. Therefore, we construct the likelihood by modelling only the count of individuals positive on both tests, denoted $y_{11\cdot}$, using a single binomial distribution:

\begin{equation*} 
y_{11\cdot} \sim \text{Binomial}(N_{S_i}, p_{11\cdot|S_i}), 
\end{equation*}

where $N_{S_i}$ is the total sample size of the subset $S$. The probability parameter $p_{11\cdot|S_i}$ represents the probability of a joint positive result conditional on inclusion in $S$, defined as the total joint probability in the full population divided by the probability of selection:

\begin{equation*} 
 p_{11\cdot|S_i} = \frac{p_{111_i} + p_{110_i}}{\mathbb{P}(S_i)}. 
\end{equation*}

\vspace{0.4cm}

\noindent \textbf{Check the negatives study}: Say that Test 1 and Test 2 are given to every participant, but only Test 2 negatives are further followed up by the gold standard. This designs might be used when the aim is to evaluate Test 1, but it is not feasible to carry out the gold standard on everyone, hence Test 2 is used as an initial imperfect reference standard. In this design, we observe four of the eight cells in Table \ref{tab: fully paired design} directly (those for participants who test negative on Test 2). However, for those testing positive on Test 2, the true disease status is unobserved. Therefore we observe only $y_{01\cdot i} = y_{011i} + y_{010i}$ and $y_{11\cdot i} = y_{111i} + y_{110i}$. This results in six observable outcomes, which we model using a 6-dimensional multinomial distribution:

\begin{align*} 
\mathbf{y}_{e_i} \sim \text{Multinomial}(N_i, \mathbf{p}_{e_i}),  
\end{align*}

where $\mathbf{y}_{e_i} = (y_{000i}, y_{100i}, y_{001i}, y_{101i}, y_{01\cdot i}, y_{11\cdot i})$ is the vector of observed counts, and \linebreak $\mathbf{p}_{e_i} = (p_{000i}, p_{100i}, p_{001i}, p_{101i}, p_{01\cdot i}, p_{11\cdot i})$ is the vector of corresponding cell probabilities. This is the design of \citet{edmonson2005} from Case Study 2.

However, as the objective of studies with this design is typically only to evaluate Test 1, we additionally note that some studies might only report 4, rather than 6 data counts, as in \citet{lindgren2016} from Case Study 2. In this study, cross-classified data were not reported at all, further collapsing the 8 cells of Table \ref{tab: fully paired design} into 4 observable counts of Test 1 vs the combined results of Test 2 and the gold standard. These four observed counts can be modelled using a 4-dimensional multinomial distribution:

\begin{equation*} 
\mathbf{y}_{l_i} \sim \text{Multinomial}(N_i, \mathbf{p}_{l_i}),\end{equation*}

where $\mathbf{y}_{l_i} =(y_{11 \cdot i} + y_{101i}, y_{01 \cdot i} + y_{001i}, y_{100i}, y_{0001})$ is the vector of observed counts, with corresponding cell probabilities $\mathbf{p}_{l_i} =(p_{11 \cdot i} + p_{101i}, p_{01 \cdot i} + p_{001i}, p_{100i}, p_{0001})$.

\subsection{Inference and Estimation}

For each of the two case studies, we modelled the data from each study according to the specific likelihoods, accounting for the varied study designs and reporting methods. The model was fitted in Stan \citep{stan_reference_manual}, software for Bayesian inference using Hamiltonian Monte Carlo. We specified vague priors for the model parameters as follows. 

Each element in the mean vector $\boldsymbol{\mu}$ was assigned a Normal(0, 5) prior. For $\mathit{\Sigma}$, standard practice in Stan is to decompose the prior into a scale vector and a matrix. Specifically, we let:

\begin{equation*}
    \mathit{\Sigma} = diag(\boldsymbol{\tau}) \times \mathit{\Omega} \times diag(\boldsymbol{\tau}),
\end{equation*}

where $\boldsymbol{\tau}$ represents the vector of between-study standard deviations, and $\mathit{\Omega}$ is the correlation matrix. We used a weakly informative Half-Normal(0, 1) prior for each element of the vector $\boldsymbol{\tau}$, following current recommendations from the \cite{stan_priors} and the NICE Technical Support Document for Evidence Synthesis of Diagnostic Test Accuracy \citep{TSD}. Note that prior information regarding the expected level of between-study heterogeneity for individual test parameters could be incorporated by assigning stronger informative priors directly to the elements of the $\boldsymbol{\tau}$ vector. For $\mathit{\Omega}$, we used an LKJ prior \citep{McElreath2020} with shape parameter 1. This shape parameter choice represents a jointly uniform prior over all valid correlation matrices, so that no specific correlation structure is favoured over another.

Stan was called in R \citep{R2025} using the `rstan' package \citep{rstan}. We used 3 chains, each with 5000 iterations, of which the first 1000 were discarded as warmup. To avoid divergent transitions, the sampler's target average acceptance probability, $adapt\_delta$, was increased to 0.97 \cite{betancourt_divergences}. We assessed convergence using the Gelman-Rubin statistic, $\hat{R}$, with convergence considered to be achieved if $\hat{R} \leq 1.01$ for each parameter. We also required the effective sample size to be at least 2500 for each parameter, which is equivalent to ensuring the Monte Carlo standard error was less than 2\% of the posterior standard deviation \citep{stan_reference_manual}.

\subsection{Model Comparison} \label{sec: model comparison and parameters}

For both Case Study 1 and Case Study 2, in addition to the the joint meta-analysis model described above, we also fitted the following for comparison:

\begin{enumerate}[label=(\alph*)]
    \item \textit{Joint meta-analysis assuming conditional independence:} In these analyses, we fitted the same model as described above, with varied likelihoods to accommodate varying designs, but assumed that the results of the two tests were conditionally independent given the disease status. This was achieved by setting all log-odds ratio parameters to zero. In this case, the joint probabilities are the simple products of the marginal sensitivities and specificities.
    \item \textit{Separate bivariate meta-analysis:} independent meta-analyses for each index test, ignoring varying study designs and treating the data as if each test were reported against a gold standard in isolation. In Case Study 1, where every participant received the gold standard in every study, this simply means analysing each test separately. This differs from the joint conditional independence approach which assumes no within-study correlations by also assuming independence between studies. However, recall that in Case Study 2, not everyone received the gold standard in every study. In this case, we meta-analysed RADT accuracy using the 4 studies where it was reported against the gold standard, and McIsaac accuracy using all the studies, but treating the reference standard as a gold standard even if it was not the case. This reflects the approach commonly taken in many systematic reviews, as in the original case study review \citep{kanagasabai2024}, but it fails to correct for potential verification bias or account for the joint data structure.
\end{enumerate}

For each model described above as well as our joint meta-analysis model from Section \ref{sec: jont MA framework}, the following parameters were monitored and summarised by posterior medians and 95\% credible intervals (CrIs):

\begin{enumerate}[label=(\alph*)]
    \item \textit{Individual test performance:} summary sensitivities ($TPR_1^s, TPR_2^s$) and specificities ($1-FPR_1^s, 1-FPR_2^s$).
    \item \textit{Conditional dependence:} the summary odds ratios, $OR_0^s$ and $OR_1^s$. Note that these are only monitored for the joint model with dependence, as in the independent case they are set to 1 by assumption, and are not estimated in separate meta-analysis.
    \item \textit{Composite rules:} we estimate the accuracy of the two index tests used in sequence, according the all positive and any positive rules. According to the all positive rule, an individual is considered to have tested positive overall if both index tests are positive, and negative otherwise, while for the any positive rule, an individual is considered to have tested positive overall if they test positive on at least one of the index tests, and negative only when both tests are negative \citep{naaktgeboren2013, fanshawe2024}. We monitor the sensitivity, specificity and the false positive rates of both of these diagnostic strategies, calculated as follows:
    
    \begin{align*}
        Sens_{all\text{ } positive} &= \text{joint TPR} \\
        Sens_{any\text{ } positive} &= TPR_1^s + TPR_2^s - \text{joint TPR} \\
        Spec_{all\text{ } positive} &= 1 - \text{joint FPR}\\
        Spec_{any\text{ } positive} &= 1 - FPR_1^s - FPR_2^s + \text{joint FPR}.
    \end{align*}
    
    Note that the sensitivity of the all positive rule is the joint TPR of the two tests, and the false positive rate of the all positive rule is the joint FPR of the two tests. 
    \item \textit{Comparative diagnostic accuracy:} we monitor the differences between summary sensitivities and specificities between the two tests, along with the probability that one test is superior to the other (in terms of sensitivity, specificity, or both).
\end{enumerate}

\section{Results}

In this section we present results for each Case Study.

\subsection{Case Study 1 Results}

Result for Case Study 1 are shown in Table \ref{tab:ultrasound_results}. 

\begin{table}[htbp]
\centering
\caption{Case Study 1: summary estimates (posterior medians and 95\% credible intervals) of the accuracy of ultrasound markers (shortened femur (F) and humerus (H)) for Down Syndrome diagnosis. Estimates are reported for the joint model with dependence, joint model assuming independence and for the separate meta-analyses of markers. Probabilities are reported as posterior means.}
\label{tab:ultrasound_results}
\small
\begin{tabular}{lccc}
\toprule
\textbf{Parameter} & \textbf{Joint Model} & \textbf{Joint Model} & \textbf{Separate} \\
 & \textbf{(Dependent)} & \textbf{(Independent)} & \textbf{Meta-Analyses} \\
\midrule
\textit{Individual Test Performance} & & & \\
Sensitivity (F) & 0.352 [0.239, 0.477] & 0.361 [0.246, 0.485] & 0.355 [0.237, 0.483] \\
Sensitivity (H) & 0.338 [0.246, 0.445] & 0.342 [0.249, 0.453] & 0.345 [0.252, 0.461] \\
Specificity (F) & 0.927 [0.895, 0.950] & 0.927 [0.895, 0.950] & 0.926 [0.896, 0.949] \\
Specificity (H) & 0.950 [0.925, 0.968] & 0.951 [0.925, 0.968] & 0.951 [0.926, 0.968] \\ \\

\textit{Correlation Parameters} & & & \\
OR0 (Non-Diseased) & 21.83 [11.42, 44.89] & 1.00 [Fixed] & --- \\
OR1 (Diseased) & 61.44 [16.24, 294.37] & 1.00 [Fixed] & --- \\ \\

\textit{Composite Rule: All Positive} & & & \\
Sensitivity (joint TPR) & 0.283 [0.194, 0.379] & 0.123 [0.070, 0.199] & 0.122 [0.075, 0.185] \\
Specificity (1 - joint FPR) & 0.974 [0.958, 0.985] & 0.996 [0.993, 0.998] & 0.996 [0.994, 0.998] \\ \\

\textit{Composite Rule: Any Positive} & & & \\
Sensitivity & 0.407 [0.316, 0.517] & 0.582 [0.462, 0.696] & 0.581 [0.479, 0.682] \\
Specificity & 0.903 [0.870, 0.928] & 0.880 [0.836, 0.913] & 0.879 [0.845, 0.908] \\ \\

\textit{Comparative Accuracy} & & & \\
Diff. in Sensitivity (F$-$H) & 0.014 [$-$0.117, 0.138] & 0.020 [$-$0.114, 0.137] & 0.010 [$-$0.155, 0.166] \\
Diff. in Specificity (F$-$H) & $-$0.023 [$-$0.054, 0.004] & $-$0.024 [$-$0.053, 0.003] & $-$0.025 [$-$0.060, 0.008] \\
Prob. F $>$ H Sensitivity & 0.587 & 0.627 & 0.550 \\
Prob. F $>$ H Specificity & 0.041 & 0.039 & 0.067 \\
Prob. F $>$ H Both & 0.016 & 0.015 & 0.015 \\
\bottomrule
\end{tabular}
\end{table}

While there is some variation in the pooled estimates of individual sensitivities and specificities, the posterior medians are close and the 95\% CrIs largely overlap across the three modelling approaches. Our joint dependent model estimates a high degree of conditional dependence between tests, with posterior median odds ratios of 21.83 for the disease-free population ($OR_0$) and 61.44 for the diseased population ($OR_1$). This suggest that the two ultrasound markers frequently result in the same diagnostic errors. Because of this high correlation, there is a clear difference in the estimated performance of composite rules when comparing the dependent model with those assuming independence. 

Under the all positive rule, the joint sensitivity is underestimated and the joint specificity is overestimated when conditional independence is incorrectly assumed. The joint sensitivity is estimated as 0.283 [0.194, 0.379] under the dependent model in contrast to 0.122 [0.075, 0.185] under the separate analysis of the tests (and similar results under the independent joint analysis.) The joint specificity is estimated as 0.974 [0.958, 0.985] with our dependent model compared to 0.996 [0.994, 0.998] from the separate analysis and almost identical results from the independent model. 

In contrast, under the any positive rule the joint sensitivity is overestimated and joint specificity  underestimated when conditional independence is incorrectly assumed. The joint sensitivity is estimated as 0.407 [0.316, 0.517] under the joint dependent model in contrast to an estimate of 0.582 [0.462, 0.696] from the joint independent model, and the estimate of joint sensitivity is 0.903 [0.870, 0.928] under the dependent model and 0.880 [0.836, 0.913] under the joint independent model. Again, the separate meta-analysis results very similar to the joint independent model. 

Estimated differences in sensitivities and specificities across the models are similar, with overlapping CrIs. While models provide slightly different levels of certainty regarding test superiority, all three models consistently confirm a high probability ($>93\%$) that the humerus is the more specific marker, while the probability that either ultrasound marker is superior in both sensitivity and specificity remains negligible ($<2\%$).

\subsection{Case Study 2 Results}

The results for Case Study 2 are presented in Table \ref{tab:mcisaac_results}.

\begin{table}[htbp]
\centering
\caption{Case Study 2: summary estimates (posterior medians and 95\% credible intervals) of the accuracy of the McIsaac rule (M) and rapid antigen detection test (R) for identifying Group A Streptococcus. Estimates are reported for the joint model with dependence, joint model with independence and for separate meta-analyses. Probabilities are reported as posterior means.}
\label{tab:mcisaac_results}
\small
\begin{tabular}{lccc}
\toprule
\textbf{Parameter} & \textbf{Joint Model} & \textbf{Joint Model} & \textbf{Separate} \\
 & \textbf{(Dependent)} & \textbf{(Independent)} & \textbf{Meta-Analyses} \\
\midrule
\textit{Individual Test Performance} & & & \\
Sensitivity (M) & 0.803 [0.637, 0.913] & 0.809 [0.640, 0.920] & 0.801 [0.637, 0.909] \\
Sensitivity (R) & 0.816 [0.593, 0.924] & 0.828 [0.593, 0.937] & 0.828 [0.576, 0.942] \\
Specificity (M) & 0.432 [0.262, 0.628] & 0.410 [0.249, 0.601] & 0.405 [0.247, 0.584] \\
Specificity (R) & 0.970 [0.923, 0.989] & 0.970 [0.926, 0.989] & 0.970 [0.921, 0.990] \\ \\

\textit{Correlation Parameters} & & & \\
OR0 (Non-Diseased) & 3.18 [0.89, 13.21] & 1.00 [Fixed] & --- \\
OR1 (Diseased) & 1.45 [0.41, 3.49] & 1.00 [Fixed] & --- \\

\\
\textit{Composite Rule: All Positive} & & & \\
Sensitivity (joint TPR) & 0.652 [0.457, 0.791] & 0.659 [0.453, 0.802] & 0.654 [0.427, 0.800] \\
Specificity (1 - joint FPR) & 0.976 [0.939, 0.992] & 0.983 [0.956, 0.994] & 0.982 [0.951, 0.994] \\ \\

\textit{Composite Rule: Any Positive} & & & \\
Sensitivity & 0.956 [0.885, 0.988] & 0.968 [0.909, 0.991] & 0.966 [0.899, 0.991] \\
Specificity & 0.425 [0.258, 0.617] & 0.397 [0.240, 0.579] & 0.391 [0.239, 0.566] \\

\\
\textit{Comparative Accuracy} & & & \\
Diff. in Sensitivity (M$-$R) & $-$0.010 [$-$0.221, 0.236] & $-$0.017 [$-$0.221, 0.252] & $-$0.026 [$-$0.222, 0.233] \\
Diff. in Specificity (M$-$R) & $-$0.534 [$-$0.702, $-$0.335] & $-$0.559 [$-$0.721, $-$0.360] & $-$0.562 [$-$0.723, $-$0.377] \\
Prob. M $>$ R Sensitivity & 0.460 & 0.432 & 0.397 \\
Prob. M $>$ R Specificity & 0.000 & 0.000 & 0.000 \\
Prob. M $>$ R Both & 0.000 & 0.000 & 0.000 \\
\bottomrule
\end{tabular}
\end{table}

We again see that for the individual test accuracies, while there is some variation in the posterior medians, the estimates are consistent across the three models. This implies that explicitly adjusting for the varying designs in Case Study 2 (verification bias, imperfect reference standard) did not substantially alter the accuracy estimates compared to the naive meta-analysis. 

For this example, the estimated dependence between the two tests is much lower than in Case Study 1. The posterior median summary odds ratio is 3.18 [0.89, 13.21] for the disease-free population and 1.45 [0.41, 3.49] for the diseased population, suggesting weak to negligible conditional dependence in both disease states. Notably, the credible intervals for both $OR_0$ and $OR_1$ include 1. 

As we would anticipate with lower levels of correlation, the differences in the estimated accuracies of the composite rules across models is less pronounced than in Case Study 1. There is little variation in the estimated accuracies of the all positive rule across modelling approaches. The estimates of joint TPR and joint FPR are similar and the 95\% CrIs largely overlap. There is a more noticeable shift in estimates under the any positive rule, where sensitivity is overestimated while specificity is underestimated when assuming conditional independence of tests. Under our joint dependent model the sensitivity and specificity are estimated as 0.956 [0.885, 0.988] and 0.425 [0.258, 0.617], compared to a sensitivity of 0.968 [0.909, 0.991] and a specificity of 0.397 [0.240, 0.579] under the joint independent model (and the second gives almost identical results with the separate analysis). However, while there is a shift in the posterior medians, the 95\% CrIs overlap. 

For the comparative accuracy of the two tests, the differences in sensitivities and specificities are similar across all models with overlapping CrIs of similar widths. While models provide slightly different levels of certainty regarding sensitivity (Prob M $>$ R ranges from 0.397 to 0.460), all three models consistently confirm with 100\% probability that the RADT is the more specific marker.

\section{Discussion}
 
In this paper, we proposed a Bayesian hierarchical modelling framework for the joint meta-analysis of two tests. Our model is designed to handle varying study and reporting designs -- for example, where full joint data on the two tests is not reported or imperfect reference standards are used -- without relying on an initial data imputation step or an assumption that the reference standard used in all studies was error-free. Moreover, it addresses the challenge of modelling conditional dependence across studies, by using log-odds ratios to parametrize the correlation and fitting a random effects structure for these across studies. We applied this framework to two case studies: first, a joint meta-analysis of two ultrasound markers (shortened humerus and shortened femur) for detecting Down syndrome; and second, a joint meta-analysis of the McIsaac clinical prediction rule and rapid antigen detection tests (RADT) for diagnosing Group A Streptococcus.

Results show that when conditional dependence is estimated to be high, as in Case Study 1, failing to account for it substantially alters the estimated accuracy of tests used in sequence. Because substantial dependence is detected in the data, results from models that ignore it are likely misleading. For example, under the all positive rule (the joint TPR of the two tests), sensitivity was underestimated at 12.3\% compared to 28.3\% in our joint model, while specificity was overestimated (96.6\% versus 97.4\%). When tests were combined under the any positive rule, the combined sensitivity was overestimated, and specificity underestimated relative to the dependent estimates. The direction of bias in our results under the two combination rules is consistent with what is mathematically dictated by the joint probability formulas. When conditional dependence was low, as in Case Study 2, this effect was less pronounced. Even so, we still observed some differences in the estimated sensitivity and specificity of the any positive rule compared to models that assume independence. Conditional independence of tests should only be assumed after careful consideration and valid justification. Otherwise, accounting for conditional dependence is essential, especially if we are interested in the accuracy of tests used in sequence.

It is important that we acknowledge that both case studies were used as illustrative examples of the proposed framework rather than to guide clinical decision making. In particular, the Down syndrome dataset originates from the 1990s and is almost certainly outdated today. While the Case Study 2 dataset is more recent, the focus of the original meta-analysis was the accuracy of the McIsaac clinical prediction rule only. Therefore, studies evaluating RADTs in isolation were not eligible for inclusion. The studies included in this data set that evaluated RADTs may not be a representative sample of all studies evaluating RADT accuracy. Furthermore, we treated various RADT brands as exchangeable, despite potential differences in their individual accuracies \citep{Cohen16}. These concerns were intentionally set aside to focus on demonstrating the framework's ability to handle complex study designs and dependence of tests within a unified model.

A key motivation of our framework was to address the limitations of methods assuming random effects distributions for other measures of conditional dependency, such as logit-transformed joint probabilities or Pearson correlations. These approaches can lead to estimation issues, because the parameters are naturally constrained by study-specific rates. We found that when full cross-classified data is not reported for all studies, the \citet{trikalinos_2014} parametrization runs into serious estimation issues. When fitting this model Case Study 1 while accounting for the varied study designs, almost all samples drawn in Stan were divergent. In contrast, while the alternative Bayesian software JAGS \citep{JAGS} appeared to run the model without errors, closer inspection revealed posterior draws where joint rates exceeded marginal rates. These imply negative multinomial probabilities, which JAGS failed to flag because the marginal studies do not use those specific cells in the likelihood. Our framework avoids these issues by using log-odds ratios to parametrize dependence. This ensures that the parameter space is always valid, resulting in superior computational stability and providing consistent estimates regardless of study designs and without needing to rely on data imputation where data is missing.

The model by \citet{Novielli2013} avoids the estimation issues associated with models using random effects on joint rates by using a stratification approach, where one test is used as a risk stratifier for the other, and test accuracies are estimated in each risk category. Overall marginal and joint test rates then can be calculated. This approach successfully overcomes estimation issues, and is especially convenient when one test is categorical, like in our Case Study 2. While for the application of our model, we chose to treat the McIsaac rule as binary with a defined positive threshold, it actually has multiple risk categories and is therefore a natural choice for the risk stratifier test in the Novielli et al. approach. However, in the case of binary tests, the choice of which test to use as the risk stratifier is often arbitrary. Since the model is not symmetrical, this choice might affect the results. For example, using data from our Case Study 1 (using only studies with full cross-classified data), where we have two binary tests, we found that switching the tests led to a significant difference in the estimated joint specificity according to the any positive rule.  

While our framework is able to handle various study and reporting designs, the specific likelihoods described here focus on those encountered in our motivating examples. This includes fully paired and partially paired designs, where participants receive one or both index tests and are followed up by the gold standard, as well as designs with alternative verification methods, where not everyone is tested on a gold standard. Other designs exist in the literature where the application of tests is based on different mechanisms that we have not explicitly modelled. For instance, we could have a check the positives designs (symmetrical to our check the negatives from Case Study 2, where only positives on the second index test are verified by gold standard) or studies using discrepant resolution designs, where the reference standard is only applied when the first two index test results disagree. In our model, we define custom likelihoods for different designs and reporting, and the same could be done for study types not detailed here. Our framework can be easily extended to include these, provided the observed cell probabilities can be expressed in terms of the multinomial probabilities defined in Table \ref{tab: fully paired design}.

Moreover, our framework can corporate single test accuracy studies, where only one of the tests is evaluated, like \citet{palla2012} in Case Study 2. However, in comparative meta-analysis, single test studies should be handled with caution. \citet{Takwoingi_2013} found that evidence derived from noncomparative studies often differs from that derived from comparative studies. When both comparative and noncomparative studies are included in the same review, the recommendation by the \citet{CochraneHandbook2023} is to conduct separate analysis and discuss any differences in results. 

There are some limitations to our proposed framework. First, extending the model to higher dimensions would be challenging. As the number of tests grows, the number of potential dependencies increases, making the log-odds ratio parametrization much more complex and harder to implement. Moreover, while it has the flexibility to accommodate varying study designs, this requires defining custom likelihoods for each design which then need to be manually coded for every new scenario. There is currently no software to handle this, making the initial setup more complex as the variety of study types increases. Developing user friendly software would make our approach more accessible and widely applicable. 

\section*{Data availability}

The case studies use data extracted from previously published studies; the extracted data are available in Tables 4 and 5 in the Appendix. The code for fitting the model to both case studies is publicly available from GitHub at \url{https://github.com/verahudak/joint_MA_two_tests}.

\section*{Appendix}

\appendix
\section{Data extracted from case studies}
\label{appendix: data for case studies}

\begin{table}[htbp]
\centering
\caption{Cross-classification of shortened femur (F) and shortened humerus (H) results among infants with Trisomy 21 (diseased) and healthy infants (disease-free). Data is grouped by study design and reporting format.}
\label{tab:ultrasound_data}
\resizebox{\textwidth}{!}{%
\begin{tabular}{l c cccc c cccc}
\toprule
 & \multicolumn{5}{c}{\textbf{Diseased ($D=1$)}} & \multicolumn{5}{c}{\textbf{Disease-Free ($D=0$)}} \\
\cmidrule(lr){2-6} \cmidrule(lr){7-11}
\textbf{Study} & $N$ & $F^+H^+$ & $F^+H^-$ & $F^-H^+$ & $F^-H^-$ & $N$ & $F^+H^+$ & $F^+H^-$ & $F^-H^+$ & $F^-H^-$ \\
\midrule
\multicolumn{11}{l}{\textit{Fully paired design with complete cross-classification}} \\
\citet{benacerraf1991} & 24 & 9 & 1 & 3 & 11 & 400 & 19 & 21 & 6 & 354 \\
\citet{benacerraf1992} & 32 & 17 & 6 & 0 & 9 & 588 & 23 & 40 & 11 & 514 \\
\citet{benacerraf1994} & 45 & 20 & 2 & 0 & 23 & 106 & 1 & 3 & 2 & 100 \\
\citet{biagiotti1994} & 27 & 10 & 3 & 0 & 14 & 500 & 31 & 29 & 29 & 411 \\
\citet{nyberg1993} & 45 & 8 & 3 & 3 & 31 & 942 & 15 & 29 & 27 & 871 \\
\\

\multicolumn{11}{l}{\textit{Fully paired design with marginal reporting only}} \\
\citet{nyberg1998} & 142 & \multicolumn{4}{c}{$F^+=30, \ H^+=27$} & 930 & \multicolumn{4}{c}{$F^+=43, \ H^+=11$} \\
\citet{rodis1991} & 11  & \multicolumn{4}{c}{$F^+=2, \ H^+=7$}   & 1470 & \multicolumn{4}{c}{$F^+=74, \ H^+=74$} \\
\\

\multicolumn{11}{l}{\textit{Fully paired with partial reporting (Cross-classification in D = 1 only)}} \\
\citet{vintzileos1996} & 22 & 4 & 1 & 6 & 11 & 493 & \multicolumn{4}{c}{$F^+=50, \ H^+=49$} \\
\\

\multicolumn{11}{l}{\textit{Partially paired with random subset (Humerus missing for some D=0)}} \\
\citet{bromley1997} & 53 & \multicolumn{4}{c}{$F^+=25, \ H^+=19$} & 177 & \multicolumn{4}{c}{$F^+=14, \ H^+=5^{\dagger} \ (n_H=149)$} \\
\citet{johnson1995}& 36 & \multicolumn{4}{c}{$F^+=15, \ H^+=8$}  & 794 & \multicolumn{4}{c}{$F^+=127, \ H^+=25^{\dagger} \ (n_H=486)$} \\
\cite{lockwood1993} & 42 & \multicolumn{4}{c}{$F^+=6, \ H^+=6$}   & 4874 & \multicolumn{4}{c}{$F^+=161, \ H^+=111^{\dagger} \ (n_H=2775)$} \\
\bottomrule
\multicolumn{11}{l}{\footnotesize $F^+$: Shortened Femur positive; $H^+$: Shortened Humerus positive.}\\
\multicolumn{11}{l}{\footnotesize $^{\dagger}$ In these studies, some Humerus results are missing completely at random, and studies only observed $n_H$ Humerus results.}\\
\end{tabular}%
}
\end{table}

\begin{table}[htbp]
\centering
\caption{Data from studies assessing the McIsaac clinical prediction rule at a threshold of 3+ and a rapid antigen detection test (RADT) for Group A Streptococcus. Counts are reported for diseased and disease-free individuals where available. For studies with alternative designs, reported marginal or restricted counts are shown.}
\label{tab:mcisaac_data}
\resizebox{\textwidth}{!}{%
\begin{tabular}{l c cccc c cccc}
\toprule
 & \multicolumn{5}{c}{\textbf{Diseased ($D=1$)}} & \multicolumn{5}{c}{\textbf{Disease-Free ($D=0$)}} \\
\cmidrule(lr){2-6} \cmidrule(lr){7-11}
\textbf{Study} & $N$ & $M^+R^+$ & $M^+R^-$ & $M^-R^+$ & $M^-R^-$ & $N$ & $M^+R^+$ & $M^+R^-$ & $M^-R^+$ & $M^-R^-$ \\
\midrule
\multicolumn{11}{l}{\textit{Fully paired design with complete cross-classification}} \\
\citet{cohen2012} & 285 & 226 & 31 & 21 & 7 & 500 & 25 & 388 & 2 & 85 \\
\citet{ezike2005}  & 154 & 59 & 5 & 85 & 5 & 209 & 3 & 80 & 1 & 125 \\
\\

\multicolumn{11}{l}{\textit{Fully paired design with marginal reporting only}} \\
\citet{abd_el_ghany2015} & 142 & \multicolumn{4}{c}{$M^+=112, \ R^+=54^{\dagger}$} & 160 & \multicolumn{4}{c}{$M^+=58, \ R^+=16^{\dagger}$} \\
\\

\multicolumn{11}{l}{\textit{Single test accuracy (McIsaac only)}} \\
\citet{palla2012} & 6 & \multicolumn{4}{c}{$M^+=6$} & 131 & \multicolumn{4}{c}{$M^+=41$} \\
\\

\multicolumn{11}{l}{\textit{Partially paired with non-random subset (McIsaac only on $D = 1$ or $R^+$)}} \\
\citet{felsenstein2014} & 58 & \multicolumn{4}{c}{$R^+=32, \ M^+_{sub}=32 \ (n_{sub}=46)$} & 303 & \multicolumn{4}{c}{$R^+=3, \ M^+_{sub}=16 \ (n_{sub}=23)$} \\
\\

\multicolumn{11}{l}{\textit{Imperfect Reference Standard (M vs R only)}} \\
\citet{nishiyama2018} & \multicolumn{10}{c}{Reported $M$ vs $R$: $M^+R^+=249, \ M^+R^-=478, \ M^-R^+=70, \ M^-R^-=184$} \\
\\

\multicolumn{11}{l}{\textit{Check the Negatives (true disease status, D, only known for $R^-$)}} \\
\citet{edmonson2005} & 1184 & -- & 52 & -- & 13 & -- & -- & 544 & -- & 191 \\
 & & \multicolumn{9}{c}{\footnotesize \textit{($R^+$ outcomes unverified, true disease status unknown. Total $R^+M^+=348, R^+M^-=36$)}} \\
\citet{lindgren2016} & \multicolumn{10}{c}{Reported $M$ vs Composite Standard$^{\ddagger}$: $M^+C^+=70, \ M^+C^-=37, \ M^-C^+=68, \ M^-C^-=145$} \\
\bottomrule
\multicolumn{11}{l}{\footnotesize $M^+$: McIsaac Positive; $R^+$: RADT Positive.}\\
\multicolumn{11}{l}{\footnotesize $^{\dagger}$ Study also reported $M$ vs $R$ contingency table: $M^+R^+=52, M^+R^-=60, M^-R^+=2, M^-R^-=28$.}\\
\multicolumn{11}{l}{\footnotesize $n_{sub}$: Total number of patients in the non-random subset who received the McIsaac score ($D=1: 46; D=0: 23$).}\\
\multicolumn{11}{l}{\footnotesize $^{\ddagger}$ Composite Standard ($C$): Positive if RADT or culture was positive.}\\
\end{tabular}%
}
\end{table}

\newpage

\section{Derivation of valid solution for $p_{111}$ and $p_{110}$}
\label{appendix: negative root}
\addcontentsline{toc}{section}{Derivation of valid solution for $p_{111}$ and $p_{110}$}

From Section \ref{sec: joint sens and spec log or}, we can express $OR_1$ in terms of $p_{111}, TPR_1$ and $TPR_2$ as follows:

\begin{align*}
    OR_1 = \frac{p_{111}(1+p_{111}-TPR_1-TPR_2)}{(TPR_1-p_{111})(TPR_2-p_{111})}.
\end{align*}

Then rearranging and solving for $p_{111}$ gives:

\begin{align*}
    p_{111} = 
\begin{cases} 
      \frac{((OR_1-1)(TPR_1+TPR_2)+1) \pm \sqrt{((OR_1-1)(TPR_1+TPR_2)+1)^2 - 4OR_1(OR_1-1)TPR_1TPR_2}}{2(OR_1-1)} & OR_1 \neq 1 \\
      TPR_1TPR_2 & OR_1=1 
\end{cases}
\end{align*}

We show that when $OR_1 \neq 1$, the negative root yields the valid solution. 

Let $A = (OR_1-1)(TPR_1+TPR_2)+1$ and $B = 4OR_1(OR_1-1)TPR_1TPR_2$. Then we have the positive root solution: 

\begin{align*}
    p_{111} = \frac{A + \sqrt{A^2-B}}{2(OR_1 - 1)}.
\end{align*}

We consider $OR_1 > 1$, $0 < OR_1 < 1$ and $OR_1 = 0$.

\subsection*{$OR_1 > 1$}

Since $p_{111}$ is the joint probability of Test 1 and Test 2 both giving positive results for a diseased individual, its value cannot exceed the probability of either single test being positive ($TPR_1$ or $TPR_2$) because the group where both events happen is a subgroup of the group where each event happens alone. Therefore we must have $p_{111} \leq min(TPR_1, TPR_2)$. We show that the positive root solution violates this constraint.

First, we have that:

\begin{align*}
    p_{111} = \frac{A + \sqrt{A^2-B}}{2(OR_1 - 1)} \geq \frac{A}{2(OR_1 - 1)} = \frac{(OR_1-1)(TPR_1+TPR_2)+1}{2(OR_1 - 1)} = \frac{TPR_1+TPR_2}{2} + \frac{1}{2(OR_1-1)}.
\end{align*}

We show that $\frac{TPR_1+TPR_2}{2} + \frac{1}{2(OR_1-1)} > min(TPR_1, TPR_2)$. Without loss of generality, we can assume that $TPR_1 \leq TPR_2$. Then we require:

\begin{align*}
    \frac{TPR_1+TPR_2}{2} + \frac{1}{2(OR_1-1)} &> TPR_1 \\
    \rightarrow TPR_1 + TPR_2 + \frac{1}{OR_1-1} &> 2TPR_1 \\
    \rightarrow TPR_1 + TPR_2 + \frac{1}{OR_1-1} - 2TPR_1 &> 0 \\
    TPR_2 - TPR_1 + \frac{1}{OR_1-1} &> 0.
\end{align*}

Since $TPR_2 - TPR_1 \geq 0 $ and $\frac{1}{OR_1-1} > 0$ when $OR_1 > 1$ , the inequality above stands for all $TPR_1, TPR_2$ and $OR_1 > 1$. Therefore we have:

\begin{align*}
    p_{111} \geq \frac{TPR_1+TPR_2}{2} + \frac{1}{2(OR_1-1)} > min(TPR_1, TPR_2),
\end{align*}

which contradicts that we must have $p_{111} \leq min(TPR_1, TPR_2)$. Hence when $OR_1 > 1$, the positive root is invalid.

\subsection*{$0 < OR_1 < 1$}

Going back to $p_{111} = \frac{A + \sqrt{A^2-B}}{2(OR_1 - 1)} = \frac{N}{D}$, because $OR_1 < 1$, we have that $D < 0$. We show that $N > 0$. We consider the following two cases: $A < 0$ and $A \geq 0$. 

\subsubsection*{$A < 0$}

We want to show that $N > 0$. We require that:

\begin{align*}
    N = A + \sqrt{A^2-B} &> 0 \\
    \rightarrow \sqrt{A^2-B} &> -A \\
    \rightarrow A^2 - B &> A^2 \\
    \rightarrow -B &> 0 \\
    \rightarrow -4OR_1(OR_1-1)TPR_1TPR_2 &> 0.
\end{align*}

Now, $TPR_1, TPR_2, OR_1 \geq 0 \rightarrow 4OR_1TPR_1TPR_2 \geq 0$, while $OR_1 < 1 \rightarrow -(OR_1 - 1) > 0$. Therefore $-B > 0$ is true always, and $N > 0$. Therefore $D < 0$ and $N > 0$, and so $p_{111} = \frac{N}{D} < 0$, which is invalid as probabilities must lie between 0 and 1, and so the positive root is invalid. 

\subsubsection*{$A \geq 0$}

When $A \geq 0$, we have that $N \geq 0$.  We have the following two cases:

\vspace{0.3cm}

\noindent $N \neq 0$

\vspace{0.3cm}

\noindent When $N \neq 0$, then $N > 0$, and so $p_{111} = \frac{N}{D} < 0$ which is an invalid probability so the positive root is invalid. 

\vspace{0.3cm}

\noindent $N = 0$

\vspace{0.3cm}

\noindent When $N = A + \sqrt{A^2 - B} = 0$, we must have $A = 0$ and $\sqrt{A^2 - B} = 0$, therefore $B = 0$. If $B = 0$, must have either $TPR_1 = 0$ or $TPR_2 = 0$. Suppose $TPR_1 = 0$. Since $A = 0$, we have that:

\begin{align*}
    A = (OR_1 - 1)(TPR_1 + TPR_2) + 1 = (OR_1 - 1)TPR_2 + 1 = 0 \\
    \rightarrow TPR_2 = \frac{1}{1 - OR_1}.
\end{align*}

We are considering $0 < OR_1 < 1$, therefore $1-OR_1$ is between 0 and 1, and $TPR_2 = \frac{1}{1-OR_1} > 1$, which is not valid as sensitivity must be between 0 and 1. Therefore the positive root is invalid. 

\subsection*{$OR_1 = 0$}

In this case, the positive root is:

\begin{align*}
    p_{111} = \frac{1-TPR_1-TPR_2+|1-TPR_1-TPR_2|}{-2}.
\end{align*}

When $1-TPR_1-TPR_2 \geq 0 \rightarrow 1 \geq TPR_1 + TPR_2$, $p_{111} = TPR_1 + TPR_2 - 1$ leads to a negative probability unless $TPR_1 + TPR_2 = 1$, in which case $p_{111} = 0$, which is valid.

When $1-TPR_1-TPR_2 < 0 $, then $p_{111} = 0$, which is valid.

Therefore the positive root gives a valid $p_{111} = 0$ only when $OR_1 = 0$ and $TPR_1 + TPR_2 \geq 1$. 

In this case ($OR_1 = 0$ and $TPR_1 + TPR_2 \geq 1$), the negative root solution is given by:

\begin{align*}
    p_{111} = \frac{1-TPR_1-TPR_2-|1-TPR_1-TPR_2|}{-2}.
\end{align*}

Then $TPR_1 + TPR_2 \geq 1 \rightarrow 1-TPR_1-TPR_2 \leq 0$. Let $|1-TPR_1-TPR_2| = \alpha$. Then we have that:

\begin{align*}
p_{111} = \frac{-\alpha - \alpha}{-2} = \alpha.
\end{align*}

Because $1-TPR_1-TPR_2 \leq 0$, $|1-TPR_1-TPR_2| = \alpha = -(1-TPR_1-TPR_2)$. Therefore $p_{111} = TPR_1 + TPR_2 - 1$, which is valid.

By the same logic, we find that the negative root solution yields valid $p_{110}$.

\bibliography{ref} 

\end{document}